\documentclass[twocolumn]{aastex62}

\newcommand{\ntargets}{827~}
\usepackage{bm}

\received{December 9, 2019}
\revised{May 7, 2020}
\accepted{May 8, 2020}

\submitjournal{AAS Journals}

\shorttitle{Planet Occurrence Around Ultracool Dwarfs}
\shortauthors{Sagear et al.}

\begin{document}

\title{Upper Limits on Planet Occurrence around Ultracool Dwarfs with K2}

\correspondingauthor{Sheila A. Sagear}
\email{ssagear@bu.edu}

\author[0000-0002-3022-6858]{Sheila A. Sagear}
\affiliation{Department of Astronomy, Boston University, 725 Commonwealth Ave., Boston, MA 02215, USA}

\author[0000-0002-4794-1591]{Julie N. Skinner}
\affiliation{Institute for Astrophysical Research, Boston University, 725 Commonwealth Ave., Boston, MA 02215, USA}

\author[0000-0002-0638-8822]{Philip S. Muirhead}
\affiliation{Department of Astronomy, Boston University, 725 Commonwealth Ave., Boston, MA 02215, USA}
\affiliation{Institute for Astrophysical Research, Boston University, 725 Commonwealth Ave., Boston, MA 02215, USA}

\begin{abstract}

The occurrence of planets orbiting ultracool dwarfs is poorly constrained.  We present results from a Guest Observer program on NASA's K2 spacecraft to search for transiting planets orbiting a sample of \ntargets ultracool dwarfs.  Having found no transiting planets in our sample, we determined an upper limit on the occurrence of planets. We simulated planets orbiting our sample for a range of orbital periods and sizes. For the simulated planets that transit their host, we injected the transit light curve into the real K2 light curves, then attempted to recover the injected planets. For a given occurrence rate, we calculated the probability of seeing no planets, and use the results to place an upper limit on  planet occurrence as a function of planet radius and orbital period.  We find that short period, mini-Neptune- and Jupiter-sized planets are rare around ultracool dwarfs, consistent with results for early- and mid-type M dwarf stars. We constrain the occurrence rate $\eta$ for planets between $0.5$ and $10$ R$_{\Earth}$ with orbital periods between 1 and 26.3 days.

\end{abstract}


\section{Introduction} \label{introduction}

Statistical results from NASA's Kepler Mission suggest a rapid increase in short-period, rocky planet occurrence with decreasing stellar mass \citep[e.g.][]{Howard2012, Gaidos2014, Dressing2015, Mulders2015}.  Short-period, rocky planet occurrence appears highest for mid-type M dwarf stars, which host 1.2 planets per star with orbital periods of less than 10 days \citep{Hardegree-Ullman2019}; however, the occurrence rate for even later type stars and brown dwarfs is poorly constrained.  Extrapolating planet occurrence rates from early and mid-type M to late-type M and cooler dwarfs, or ultracool dwarfs, we expect that they host even more short-period rocky planets.  The discovery of seven planets transiting the M8 star TRAPPIST-1 \citep{Gillon2016, Gillon_2017} reinforces this expectation.  However, the intrinsic faintness of ultracool dwarfs has historically limited large, statistical investigations into their planet occurrence rates.

\citet[][]{He2017} searched for transiting planets in the photometric timeseries of 44 brown dwarfs obtained for the Weather on Other Worlds program \citep{Metchev2015}, which used NASA's Spitzer Space Telescope to study variability in L and T dwarfs.  Finding no new exoplanets, they determined the occurrence rate of planets with with orbital periods less than 1.28 days and radii between 0.75 and 3.25 R$_{\Earth}$ to be $\eta$ $<$ 67\%. \citet{Demory2016} searched for transiting planets around 189 late M dwarfs observed by K2 and found no new planets. They show that TRAPPIST-1-like planets are able to be recovered in 10\% of their sample, but "inflated" TRAPPIST-1-like planets that resemble super-earths that may still be rocky (1.5-2.5 $R_{\Earth}$) yield a higher recovery rate of up to 70\%.
 
Motivated by the lack of discoveries coupled with the expectation of short-period rocky planets, we executed a Guest Observer program on NASA's K2 Mission to search for exoplanets transiting ultracool dwarfs.  The K2 Mission repurposed the original {\it Kepler} spacecraft for continued operations using only two reaction wheels \citep[][]{Howell2014}.  Unlike the primary Kepler Mission, which observed a single 100-square-degree sector of the sky for nearly four years, the K2 Mission observed multiple 100-square-degree sectors of the sky spread across the ecliptic, each for approximately 80 days at a time.  The greater sky coverage provided the opportunity to search hundreds of relatively bright, nearby ultracool dwarfs for short-period transiting planets. 

In this paper, we present results from our K2 Guest Observer program.  We searched for transiting planets in our sample of ultracool dwarf light curves.  Similar to previous work, we did not discover new transiting planets. However, we used the non-detections to analyze the planet detection efficiency of \textit{Kepler} in this regime and place upper limits on planet occurrence as a function of planet radius and orbital period.  We used transit injection and recovery simulations to ascertain the likelihood of our null result, as a function of planet radius, orbital period and planet occurrence.

In Section 2, we discuss the sample of K2 ultracool dwarfs.  In Section 3, we outline the data reduction and search for transiting exoplanets.  In Section 4, we detail techniques used for transit injection and recovery.  In Section 5, we present our constraints on planet occurrence rate, and in Section 6 we conclude with a discussion of our findings.

\section{The Sample}

\begin{figure}
    \centering
    \includegraphics[width=3.4in]{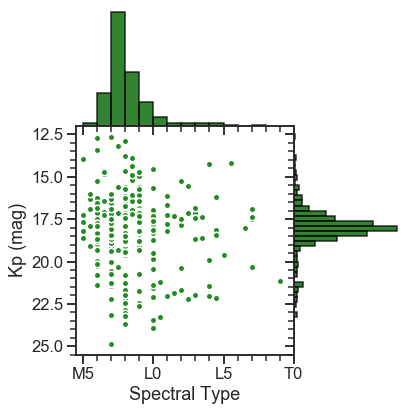}
    \caption{Spectral types and Kepler magnitudes for the \ntargets ultracool dwarfs in our sample.  Most of the objects are mid-to-late M dwarfs, with $\simeq$75 L dwarfs.} 
    \label{fig:specmag}
\end{figure}

To identify ultracool dwarfs in the K2 fields, we searched the BOSS (Baryon Oscillation Sky Survey) Ultracool Dwarf catalog (BUD; \citealt{Schmidt2015}), which contains 11,820 M7-L8 dwarfs characterized through spectroscopy.  This yielded 680 objects observable by K2 and makes up the majority of our sample.  We additionally selected spectroscopically confirmed ultracool dwarfs later than spectral type M5 from the online repositories\footnote{\href{https://jgagneastro.com/list-of-m6-m9-dwarfs/}{List of M6-M9 Dwarfs}; \href{https://jgagneastro.com/list-of-ultracool-dwarfs/}{List of Ultracool Dwarfs}} maintained by J. Gagn{\'e} \citep{Gagne2016a, Gagne2016b}\footnote{These repositories were originally built from ultracool dwarfs listed in \citet{Mace2014}, the \href{http://spider.ipac.caltech.edu/staff/davy/ARCHIVE/index.shtml}{DwarfArchives.org} catalog \citep{Gelino2016}, and \citet{Dupuy2012}.}, which yielded 101 objects in our sample.  We identified 20 objects from \citet{Skrzypek2016}, who used the cross-section of SDSS (Sloan Digital Sky Survey) and WISE (Wide-Field Infrared Survey Explorer) with UKIDSS (UKIRT Infrared Deep Sky Survey) to identify ultracool dwarfs.  In addition, 26 of the targets were identified using colors and proper motions from the cross-match of 2MASS and ALLWISE \citep{Schneider2016}.  We used the python package \texttt{k2fov} to identify which objects were within the field of view in a particular K2 campaign.  In total, we identified \ntargets late-type M dwarfs and early-type L dwarfs bright enough to be observed as a part of K2.  The sample used in our analysis contains data from campaigns 5, 6, 11, 12, 13, 14, 15, 16, 17, and 18.  Of the \ntargets objects in our sample, 781 had spectral types derived from spectroscopy.  For the 20 objects from \citet{Skrzypek2016}, we used their calculated \textit{photo-type} \citep{Skrzypek2015} from their catalog. For the remaining 26 objects, we assigned spectral types from 2MASS and ALLWISE photometry using the $J-W2$ relation from \citet{Rodriguez2013}. Figure \ref{fig:specmag} shows spectral types and Kepler magnitudes for the observed targets.

\begin{deluxetable*}{cDDDccr}
\tablenum{1}
\tablecaption{K2 Ultracool Dwarf Sample\label{sample}}
\tablewidth{0pt}
\tablehead{
\colhead{EPIC ID} & \twocolhead{$\alpha$ (deg)} & \twocolhead{$\delta$ (deg)} & \twocolhead{Kp (mag)} & \colhead{Spectral Type} & \colhead{K2 Campaign} & \colhead{\textit{Gaia} DR2 Source ID} \\
}
\decimals
\startdata
248044306 & 67.53018 & 26.13911 & 16.41 & M8.5 & 13 & 151296579553731456\\
248018652 & 67.73828 & 25.94430 & 15.90 & M8 & 13 & 151102790629500288\\
247051861 & 68.06696 & 18.21287 & 12.69 & M6 & 13 & 3314309890186259712\\
251456966 & 68.89929 & 21.25248 & 18.71 & L0 & 13 & \\
247581233 & 68.96430 & 22.81999 & 16.93 & M8.5 & 13 & 145209339585095424\\
251456967 & 74.83841 & 15.68312 & 17.59 & L0 & 13 & 3393271558253207680\\
246711015 & 75.55606 & 14.71022 & 15.63 & L0 & 13 & 3392546632197477248\\
211936497 & 123.07013 & 19.27239 & 18.07 & M7 & 18 & 669516407093922432\\
211783664 & 123.07365 & 17.08431 & 18.25 & M7 & 18 & 656647826080870016\\
211541204 & 123.13754 & 13.71910 & 18.05 & M7 & 18 & 650624465159332352\\
\enddata
\tablecomments{The first 10 entries are shown here.  The full version of this table with all \ntargets entries is available in the online version.}
\end{deluxetable*}

We provide the \textit{Gaia} DR2 Source IDs for our sample \citep{Gaia2016,Gaia2018b}. To identify \textit{Gaia} DR2 source IDs, we first cross-matched our sample and the MLSDSS catalog \citep{Kiman2019}, finding 730 matches.  687 of these matches had listed \textit{Gaia} DR2 Source IDs.  The remaining objects were cross-matched with the full \textit{Gaia} DR2 catalog using a 5" search radius.  Proper motions and observation dates from SDSS were used to propagate coordinates to the \textit{Gaia} DR2 epoch (2015.5) where available.  In the small number of cases where multiple objects fell within our search radius, we visually inspected archival imaging data to distinguish between the objects using the Aladin Sky Atlas \citep{Bonnarel2000,Boch2014}. This resulted in 58 additional matches to the \textit{Gaia} DR2 catalog.  Of the remaining 82 objects, six had high proper motions listed in the K2 Ecliptic Plane Input Catalog \citep{Huber2016}.  We further inspected archival imaging data for these six objects and found three of these to have matches in the \textit{Gaia} DR2 catalog.

Of the \ntargets targets in our sample, 248 were observed in multiple K2 campaigns. There were 211 targets observed in two campaigns and 35 targets observed in three campaigns. The same target was often observed multiple times in campaigns 5, 16, and 18, or campaigns 6 and 17.  A portion of the sample is shown in Table \ref{sample}; the full version of Table \ref{sample} is available online.

\section{Data Analysis and Planet Search}

We searched for planets around each of our \ntargets targets. For some targets, we searched for planets using more than one light curve if they were observed more than once in different campaigns. In order to increase the probability of detecting a planet, we stitched multiple observations together to create a longer light curve, preserving the timestamp of the observations. We searched these stitched light curves for planets in the same way as we search single campaign light curves. We searched a total of 1040 light curves.

We used K2 light curves reduced using the self flat-fielding method (SFF) developed by \citet{Vanderburg2014}, which correlates flux variations with the pointing variations caused by K2's spacecraft motion and removes the correlation. The flux values we use are taken from the aperture that provides the best photometric precision. The photometric precision is assessed by the the median value of a running standard deviation over bins of 13 observations, a similar metric to the Combined Differential Photometric Precision (CDPP), over a six-hour window \citep{Vanderburg2014}. We compared the following planet search and injection and recovery test using K2SFF light curves and K2 light curves reduced using the EVEREST pipeline \citep{Luger_2016}, and found no significant difference in our results.

We downloaded the SFF-reduced light curves from the Mikulski Archive for Space Telescopes (MAST).  The data described here may be obtained from the MAST archive. Due to the intrinsic faintness at optical wavelengths, many light curves show significant noise. Many objects also show evidence of rapid stellar rotation, which produces a periodic signal due to starspots rotating into and out of view, and stellar flares, which produce a sudden increase in brightness. A full analysis of the rotation and flaring activity for our sample is left as future work.

We normalized and sigma clipped the SFF-reduced light curves by discarding all points that fall farther than three sigma from the mean and replacing them with the mean value of all data points. We then detrended each light curve using a sliding window median filter with \texttt{untrendy}. This median filter creates a trend line from the median value of a sliding window of 10 points. We used \texttt{untrendy} because it is tuned to find long-term trends in Kepler data. We found that \texttt{untrendy} performed well throughout the light curves, including at the edges.

We divided each normalized light curve by its trend line. We searched for transits in the processed light curves using the python implementation \texttt{python-bls}\footnote{https://github.com/dfm/python-bls} of a box-fitting least-squares (BLS) method \citep{Kovacs2002} to create a preliminary period and radius estimate. We searched for planets with periods between 1 and at most 26.3 days using 10,000 frequency bins and transit durations ranging from $0.1\%$ to $30\%$ of the total observation time.

We did not detect any transit signals around the \ntargets ultracool dwarfs that had a high enough confidence to be considered a threshold crossing event. The rest of this work focuses on the significance of this null result.

\section{Injection and Recovery Tests}

To better understand K2's sensitivity to planets around ultracool dwarfs, we simulated planets and attempted to recover them to measure the transit detection efficiency.  We injected 10,000 randomly chosen synthetic planets into each of our 1040 K2 light curves and attempted to recover them, calculating the fraction of transits recovered in specific bins of period and radius. We used \texttt{ktransit} \citep{Barclay2015}, a transit simulation package based on the \citet{Mandel2002} limb-darkened transit model, to inject planets into the K2 light curves.  

We assumed a stellar mass $M_{s}$ of $0.1 M_{\odot}$ and a stellar radius $R_{s}$ of $0.1 R_{\odot}$, a reasonable estimation due to the narrow range of our sample's spectral types (late-type M to L) and the corresponding narrow range of radii \citep{Kirkpatrick2005}.

To choose planet parameters to inject, we randomly drew a planet radius from a uniform distribution between 0.05 and 1.0 stellar radii (about 0.5 to 11 $R_{\Earth}$). We randomly drew an inclination, assuming the orbital axis of each system may lie in any direction with equal likelihood. We allowed injections of planets with inclinations that do not produce a transit to account for the geometric transit probability when calculating an upper limit to planet occurrence. For this analysis, we considered only one transiting planet in the system.

\begin{figure}
    \centering
    \includegraphics[width=3.4in]{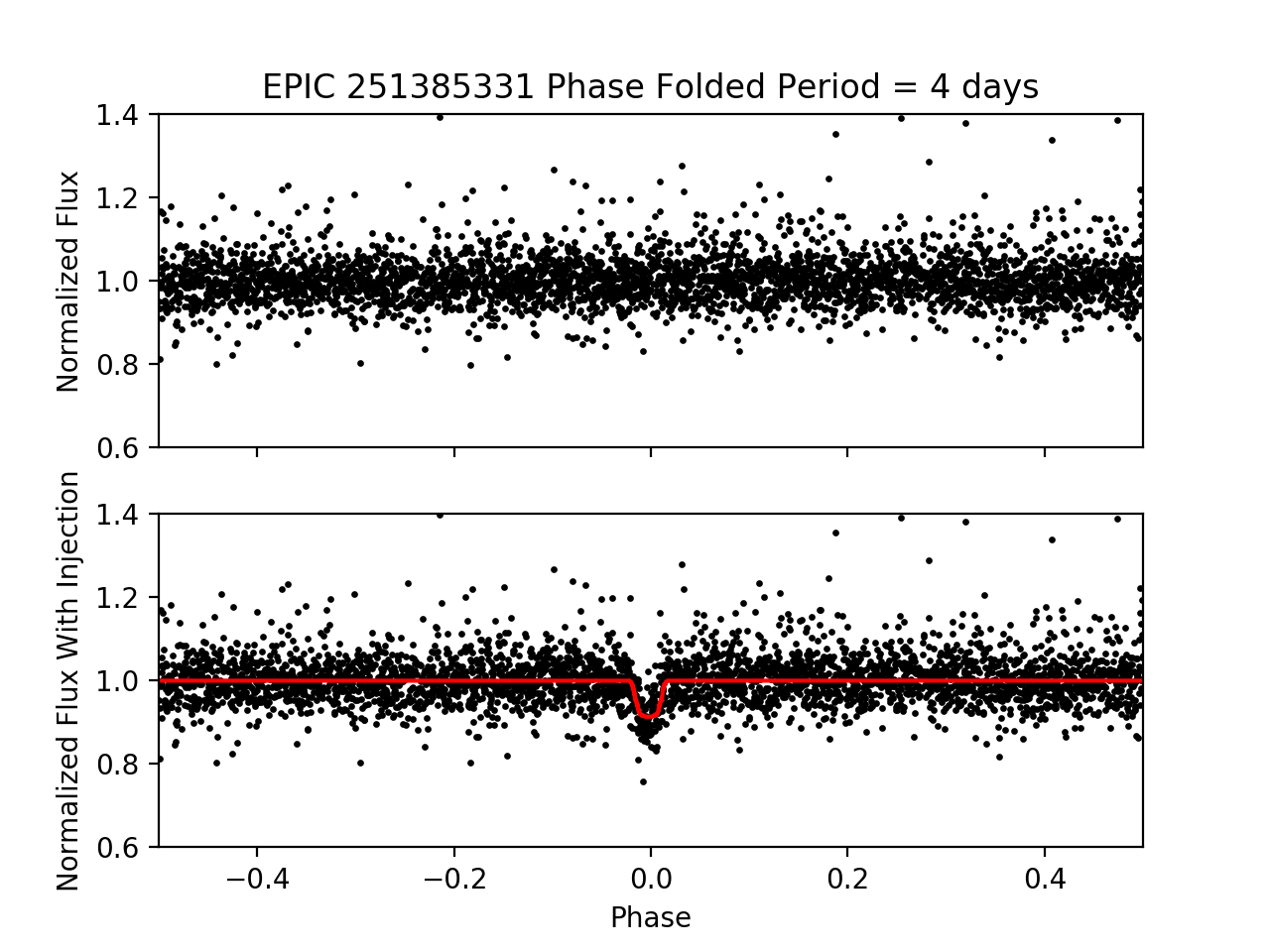}
    \caption{\textit{Top:} The K2 light curve for EPIC 251385331, detrended and phase-folded over a period of four days. \textit{Bottom:} The phase-folded and detrended light curve after injecting a synthetic planet with 0.3 stellar radii and period of four days. The red line shows the Levenberg-Marquardt fit.} 
    \label{fig:injection}
\end{figure}

As part of our detection criteria, we required that an injected planet transit at least three times. K2 observed a single patch of sky for approximately 80 days, so we only injected orbital periods between 1.0 and 26.3 days. For objects observed in campaigns that were shorter than 80 days (e.g. campaign 18), we injected and recovered planets from a period range between 1 day and one-third the observation time. We weighted the results according to the number of light curves subject to this rule. We randomly chose a value within the orbital period limits from a logarithmic distribution. The final synthetic transit light curves we created have no noise or artifacts. We multiplied the synthetic transit light curves by the SFF-reduced K2 light curves, as shown in Figure \ref{fig:injection}. 

We injected planets into the stitched, multi-campaign light curves the same way as we injected planets in single campaign light curves. When we stitched data from multiple campaigns together, we preserved the time stamps between subsequent observations so that when we injected a planet, the phase of the transit was in agreement with the observation time. 

To recover our injected planets, we used the same BLS procedure we used in our original search for planets. We then refined our transit fit using the \texttt{ktransit} implementation of Levenberg-Marquardt minimization. We used the BLS period with the highest power, the square root of the BLS transit depth, and an impact parameter of 0.0 as a starting point for the Levenberg-Marquardt fit.

For all objects with injected planets, we assumed a star with the same quadratic limb darkening model injected into all light curves. We considered a planet to be recovered if the recovered orbital period matches the injected orbital period within $5\%$, and if the recovered transit depth matches the injected transit depth within $25\%$. These thresholds allowed for transit models to reasonably match injected planets and did not produce false positives. We searched for one transiting planet at a time in each light curve.

We calculated an impact parameter $b$ for each planet from an inclination $i$ randomly chosen from a $sin(i)$ distribution using

\begin{equation}
b = \sqrt[\scriptstyle 3]{\frac{GMP^{2}}{4\pi^{2}}}\frac{\cos(i)}{R_{s}}
\end{equation}

If no part of the injected planet transits the star, or when the impact parameter is greater than 

\begin{equation}
b = 1 + \frac{R_p}{R_s}
\label{eq:impact}
\end{equation}

we did not carry on with a BLS search and transit fitting, and automatically marked the planet as unrecovered. This accounts for up to $95\%$ of all injected planets.

For each planet injection and recovery, we recorded the injected orbital period, radius, and inclination, and whether the planet was successfully recovered.

\begin{figure}
    \centering
    \includegraphics[width=3.4in]{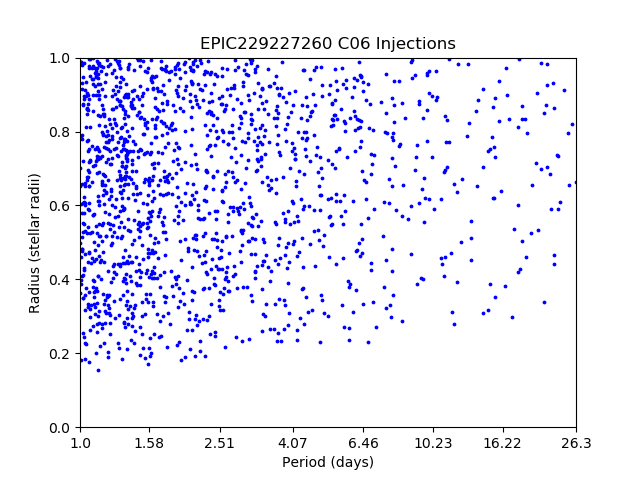}
    \caption{Injected and recovered planets around EPIC 229227260 in parameter space (orbital period vs. radius). 50,000 planets were injected with varying inclinations, about 5\% of those planets transit the star, and 60\% of those planets that transit are recovered. The division in parameter space between detectable and undetectable planets shows that most planets below 0.2 stellar radii are undetectable.}
    \label{fig:scatter}
\end{figure}

\begin{figure}
    \centering
    \includegraphics[width=3.4in]{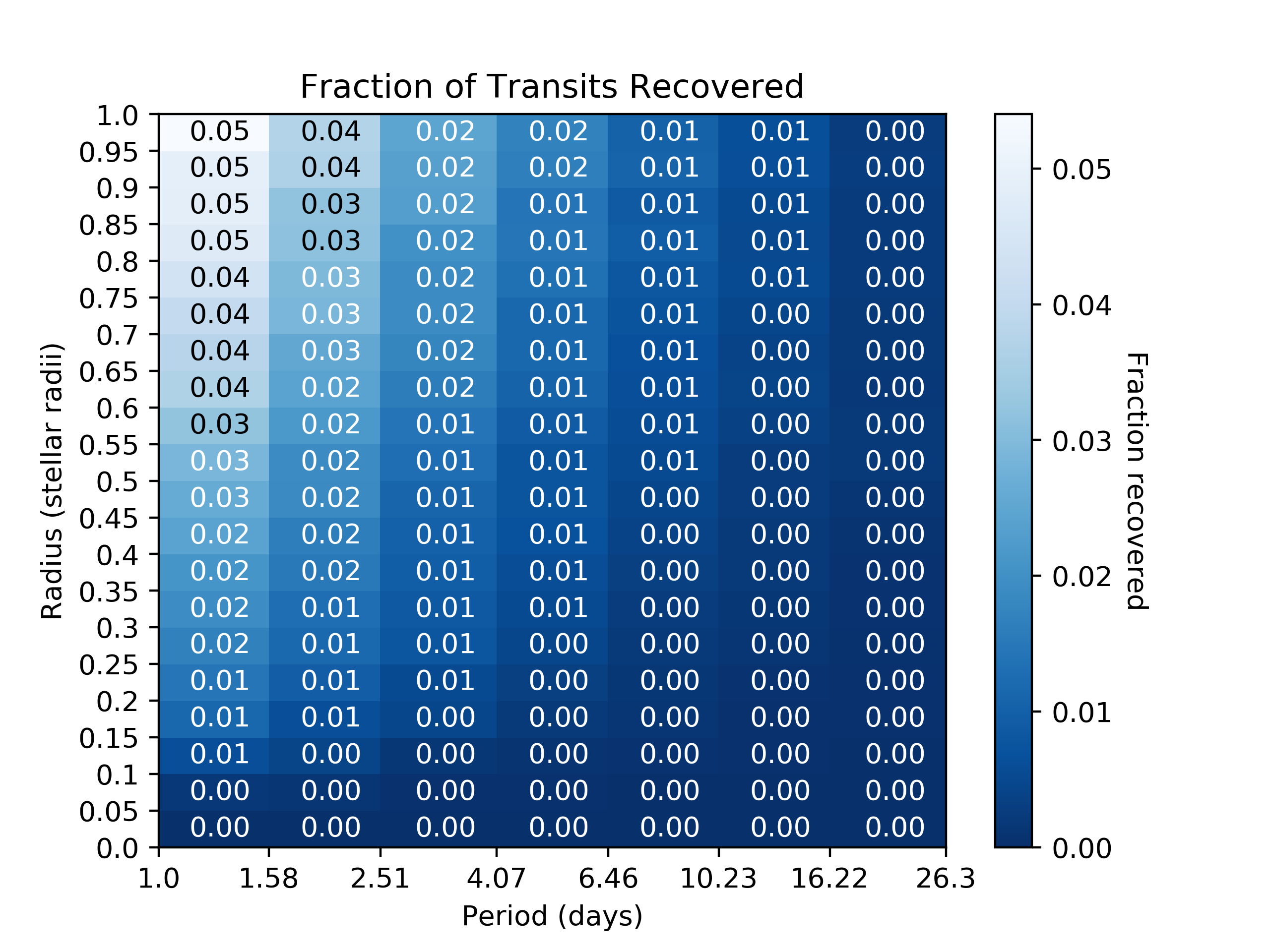}
    \caption{The fraction of planets we are able to detect after injecting 10,000 planets into each star in our sample. The detection efficiency is less than 0.05 for most bins because we vary the inclination for injected planets from 0 to 90 degrees, and most planets do not transit.}
    \label{fig:rectransits}
\end{figure}

The planet injection and recovery experiments show the threshold of detectability between planets that are generally recoverable and unrecoverable. Figure \ref{fig:scatter} shows this threshold for the K2 light curve of EPIC 229227260. We inject 50,000 random synthetic planets one by one into the light curve for this target, following the method above, and attempt to recover them. Planets between 0.15 and 0.25 stellar radii (1.5 to 2.5 $R_{\Earth}$) are recovered about $1\%$ of the time depending on the orbital period and impact parameter. Smaller impact parameters (necessarily less than $1 + \frac{R_{p}}{R_{s}}$, as in Equation \ref{eq:impact}) and shorter orbital periods increase the likelihood of detectability. We are most sensitive to planets with orbital periods between 1.0 and 6.5 days and with radii larger than 0.15 ${R_s}$ (1.5 $R_{\Earth}$), where $5\%$ of injected planets were recovered around this object. For periods longer than 6.5 days, we recovered only $0.09\%$ of injected planets around this object.
The detectability threshold changes based on target brightness and signal-to-noise ratio of the light curve. The composite percentage of planets recovered, which we call the detection efficiency, is used to calculate the upper limit on planet occurrence.  Figure \ref{fig:rectransits} shows the fraction of transits we recover as a function of orbital period and $\frac{R_{p}}{R_{s}}$.  For each bin of planet period and radius, we calculate the average detection efficiency, or percent of planets recovered, over 10,000 injections into the \ntargets K2 targets.

\section{Planet Occurrence}

We define $\eta$ as the rate of occurrence of a planet within an orbital period and radius bin. The occurrence rate may range from 0 to 1. We make the underlying assumption that there is not more than one planet in each bin. In bins where we recover no planets, we have no information on the true occurrence rate, so the upper limit on $\eta$ will be 1.0.

We calculated the probability distribution of detecting no planets around $n=827$ ultracool dwarfs based on the detection efficiency and possible occurrence rates of planets. Each calculation is made for a planet within a specific orbital period and radius bin. Considering a specific orbital period and radius range, we multiplied the detection efficiency ($d_{i}$) for this range by every possible planet occurrence rate (between 0 to 1). The probability of detecting no planets in a specific bin is shown in Equation \ref{pnulleq}.

\begin{equation}
P_{null} = \prod_{i=1}^{n} 1-\eta d_{i}
\label{pnulleq}
\end{equation}

This product gives us the probability of not seeing a planet around an ultracool dwarf as a function of planet occurrence rate $\eta$ from 0 to 1. The value of $P_{null}$ defines the confidence level for the upper limit of $\eta$. The relationship between $P_{null}$ and $\eta$ can effectively be treated as a probability distribution of $\eta$.

\begin{figure}
    \centering
    \includegraphics[width=3.4in]{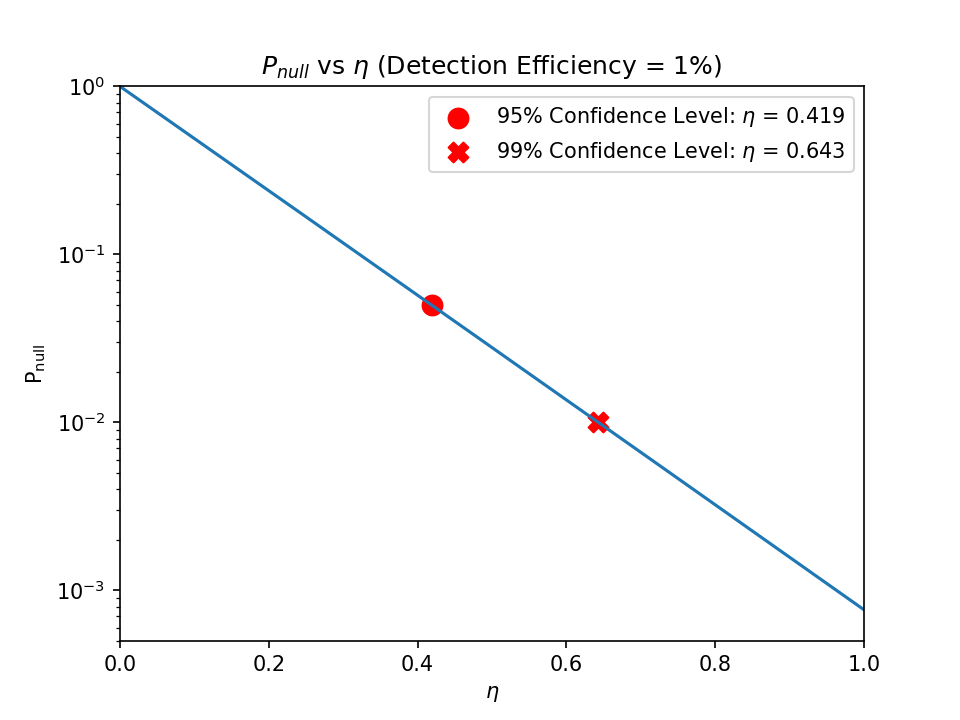}
    \caption{$P_{null}$ is the probability of seeing no planets as a function of true planet occurrence rate. This plot describes $P_{null}$ for planets with periods between 2.51 to 4.07 days and $R_p/R_s$ between 0.3 and 0.35. The red circle shows the upper limit of $\eta$ with 95\% confidence, and the red cross shows the upper limit of $\eta$ with 99\% confidence.}
    \label{fig:pnullline}
\end{figure}

We first set the probability of seeing zero planets around \ntargets ultracool dwarfs to less than $5\%$, and we calculated the upper limit of true planet occurrence $\eta$ for each binned planet type. In Figure \ref{fig:pnullline}, a probability distribution of $\eta$ for planets with orbital period between 2.51 and 4.07 days and radii from 0.3 to 0.35 stellar radii (3.0 to 3.5 $R_{\Earth}$), we determine that setting $P_{null}$ to less than $5\%$ yields an upper limit for $\eta$ of 0.419. This means that no more than $41.9\%$ of ultracool dwarfs host this type of planet. We calculated in a similar way the upper limit for $\eta$ where $P_{null}$ is less than $1\%$, which yields an upper limit for $\eta$ of 0.643.

\begin{figure}
    \centering
    \includegraphics[width=3.4in]{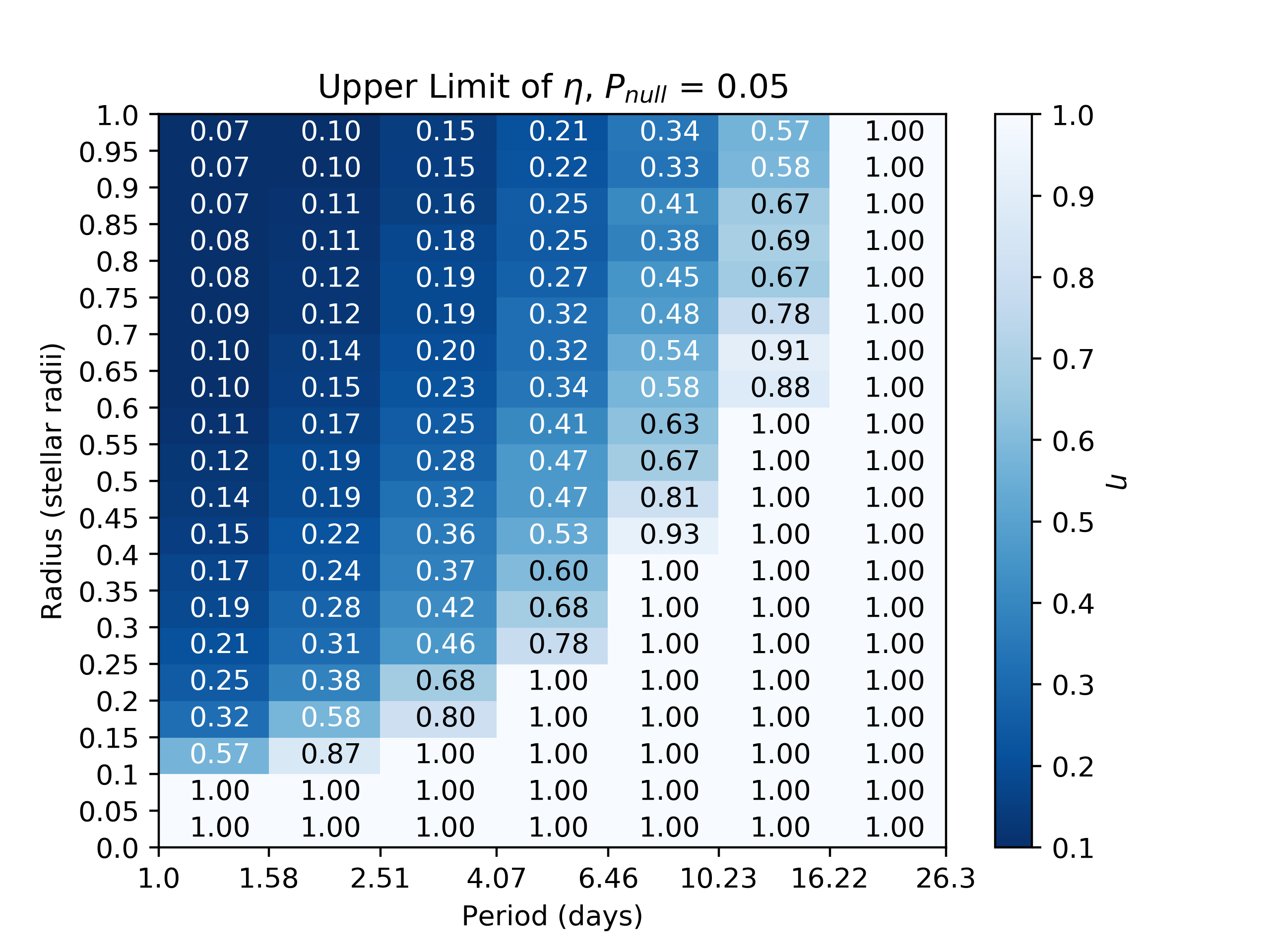}
    \caption{Upper limits on occurrence for each planet bin, with $P_{null} < 5\%$.}
    \label{fig:pnullheatmap5}
\end{figure}

\begin{figure}
    \centering
    \includegraphics[width=3.4in]{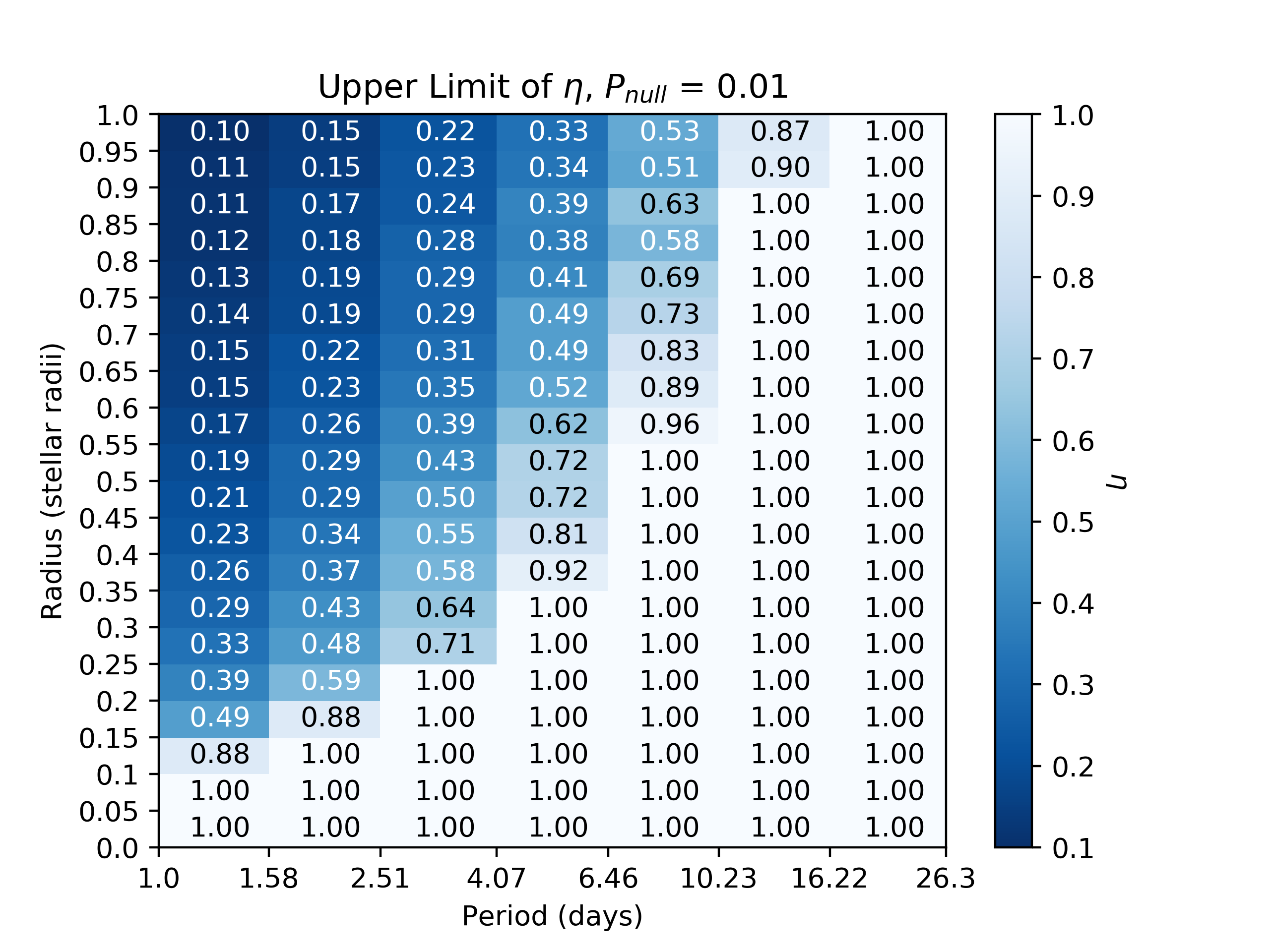}
    \caption{Upper limits on occurrence for each planet bin, with $P_{null} < 1\%$.}
    \label{fig:pnullheatmap1}
\end{figure}

We calculated an upper limit for $\eta$ in this way for each planet type according to the bins we used in Figure \ref{fig:rectransits}. Figure \ref{fig:pnullheatmap5} shows the value of the upper limit of $\eta$ for each planet type with $P_{null} < 5\%$. For short period (less than 4 days), large (greater than 0.2 stellar radii) planets that have a higher detection efficiency, $\eta \simeq$ 0.1. For planets we are not able to consistently detect, values of $\eta$ range between 0.8 and 1 (i.e. planet occurrence is not well constrained in this regime). Figure \ref{fig:pnullheatmap1} shows the upper limit on $\eta$ for each planet type with $P_{null} < 1\%$.

\section{Discussion}

We are more likely to detect short-period, large planets than long-period, small planets, as expected. The threshold for detectability is different for each target, but the smallest planets we are able to recover are super-earths, which have radii of 0.1 to 0.2 stellar radii.

\subsection{Planet Formation}
Our results are consistent with constraints on planet formation in protoplanetary disks. \citet{Rilinger_2019} measured the disk masses and radii of the protoplanetary disks around two brown dwarfs with spectral type M7. They found that one disk only contains enough material to form earth-mass planets and smaller, and one disk does not contain enough material to form planets at all. This implies that planets more massive than 1 $M_{\Earth}$ may be rare around brown dwarfs. \citet{PayneLodato2007} also determined that for a brown dwarf with mass $0.05 M_{\odot}$, planet formation following the sequential core accretion model \citep{POLLACK199662} produces a maximum planetary mass of approximately 5 $M_{\Earth}$. Our results are consistent with these findings, as the upper limit of $\eta$ is most tightly constrained for hot mini-Neptunes and Jupiter-sized planets, suggesting that large planets are rare around ultracool dwarfs.

\subsection{Comparison to \citet{He2017}}

\citet{He2017} performed a planet search on a sample of 44 brown dwarfs (L3 to T8) observed by the Spitzer Space Telescope, which similarly returned a null result. For planets with periods less than 1.28 days and radii between 0.75 and 3.25 earth radii, they determined a $95\%$ confident upper limit of $\eta < 67 \pm 1\%$. For periods less than 1.28 days and radii between 0.75 and 1.25 earth radii, they place a $95\%$ confident upper limit of $\eta < 87 \pm 3\%$.

We compare these results with ours for short-period, small planets. For planets with periods between 1 and 1.58 days and radii between 0.1 and 0.35 stellar radii (about 1.0 to 3.5 earth radii), we find that when $P_{null} < 5\%$ (equivalent to a $95\%$ confidence) we place the upper limit of $\eta$ between $57\%$ and $19\%$, depending on the planet bins. The upper limit of $\eta$ for those planets between 0.1 and 0.15 stellar radii is $57\%$.

With a sample size of \ntargets ultracool dwarfs, we are able to place a tighter constraint on $\eta$ than \citet{He2017}. However, by utilizing data from Spitzer, they are able to include objects at later spectral types.

\subsection{Comparison to \citet{Demory2016}}

\citet{Demory2016} measured transit detection efficiency for ``inflated" TRAPPIST-1b-like planets around mid- to late-M dwarfs, i.e. planets with short periods (less than three days) and up to 0.25 Jupiter radii (mini-Neptune sized). They determined that K2 is sensitive to close in mini-Neptunes, and they expect to recover $71\%$ of these planets if they transit. However, their planet search yielded no detections. They determined that hot mini-Neptunes are rare around mid- to late-M dwarfs, as they are around early- to mid-type M dwarfs \citep{Dressing2015}. 

We also find that hot mini-Neptunes are rare around late-type M dwarfs and L dwarfs. For the hot mini-Neptunes we injected, planets with periods from 1 to 4.07 days and radius between 0.2 and 0.45 $R_{s}$, we measured an average detection efficiency of $1.4\%$ over all inclinations. Taking into account a transit probability of $5.4\%$, we expected to detect $26\%$ of hot mini-Neptunes that transit, but we did not detect any of these planets. We place the upper limit of $\eta$ for hot mini-Neptunes between $58\%$ and $25\%$, suggesting that this type of planet is rare around ultracool dwarfs as well as M dwarfs at earlier spectral types.

\subsection{TRAPPIST-1}
One of the targets in our sample, EPIC 246199087 (TRAPPIST-1), is a late-type M dwarf known to host seven Earth-sized planets \citep{Gillon_2017}. We are sensitive to planets as small as 0.1 to 0.15 stellar radii (about 1 to 1.5 $R_{\Earth}$), where we detected 20\% of transiting planets (Figure \ref{fig:rectransits}). We did not detect planets around TRAPPIST-1 during the planet search stage using a BLS search and Levenberg-Marquardt fitting. However, we are able to recover planets around TRAPPIST-1 if we provide initial estimates of the TRAPPIST-1 planets' orbital periods and sizes.

\section{Summary}

We searched for transiting planets in a sample of \ntargets ultracool dwarfs observed by K2 in long-cadence mode.  The majority of our sample has spectral types between M6 to L5 determined from spectroscopy, with a handful of objects at later types.  We found no transit events that met our detection criteria and thus use this result to further investigate K2's sensitivity to transiting planets around ultracool dwarfs. 

We performed injection and recovery tests and determined planet detection efficiencies for \ntargets ultracool dwarfs.  The detection efficiencies were calculated using the BLS method and Levenberg-Marquardt optimization. Using these detection efficiencies, we determined upper limits on planet occurrence for planets between 0.05 and 1 stellar radii (about 0.5 to 10 earth radii) with periods between 1.0 and 26.3 days. We find that short period, gaseous planets are rare around ultracool dwarfs.

We note that, during the revision of this manuscript, \citet[][]{Sestovic2020} published an independent investigation into the occurrence of planets orbiting ultracool dwarfs using data from the K2 spacecraft.  Their investigation included many of the targets that were observed as part of our K2 Guest Observer Program.

In the future, it will be useful to determine detection efficiencies for multiple planets around ultracool dwarfs, since these objects have been previously shown to host groups of short-period planets in aligned systems: so-called compact multiples \citep{Ballard_2016}. Similar work with TESS observations may yield better constraints on $\eta$ for planets smaller than 0.2 stellar radii with long periods, due to TESS's redder bandpass \citep{Ricker2014}. Red-optical ground-based transit surveys \citep[e.g.][]{Gillon_2017, Burdanov_2018}, and infrared ground-based transit surveys \citep[e.g.][]{Tamburo2019}, may also yield better constraints on $\eta$ because of their sensitivity to small planets around ultracool dwarfs.

\acknowledgements

This paper includes data collected by the K2 mission under Guest Observer Programs GO6037, GO7037, GO11022, GO12022, GO13022, GO14012, GO15012, GO16012, GO17025 and GO18025. Funding for the K2 mission is provided by the NASA Science Mission directorate.

The authors acknowledge support from the NASA K2 Guest Observer Cycle 4 and K2 Guest Observer Cycle 5 programs under Grant Nos. NNX17AE91G and 80NSSC18K0294 issued through the Science Mission Directorate. The authors thank Aurora Kesseli for providing code to generate planet inclinations.  The authors also thank Adam Schneider and Michael Cushing for useful conversations and for providing early access to the ALLWISE catalog of high proper motion objects.  This research has also benefited from the M, L, T, and Y dwarf compendium housed at \href{http://spider.ipac.caltech.edu/staff/davy/ARCHIVE/index.shtml}{DwarfArchives.org}. This research has made use of ``Aladin sky atlas" developed at CDS, Strasbourg Observatory, France

S.S. acknowledges support from the Undergraduate Research Opportunities Program at Boston University and the Clare Boothe Luce Scholar Award.

The authors thank the anonymous referee for comments that improved this manuscript.

All of the data presented in this paper were obtained from the Mikulski Archive for Space Telescopes (MAST). STScI is operated by the Association of Universities for Research in Astronomy, Inc., under NASA contract NAS5-26555. Support for MAST for non-HST data is provided by the NASA Office of Space Science via grant NNX13AC07G and by other grants and contracts. 

\facilities{K2}

\software{\texttt{ktransit} \citep{Barclay2015}; \texttt{python-bls}; \texttt{k2fov} \citep{Mullally2016}; \texttt{astropy} \citep{astropy2013,astropy2018}; \texttt{Matplotlib} \citep{Hunter2007}}

\bibliographystyle{aasjournal}
\bibliography{bibfile}

\end{document}